\documentclass[manuscript]{aastex}
\shorttitle{Test of WEP with the Crab Pulsar}
\shortauthors{Zhang \& Gong}

\begin{document}


\title{Test of Weak Equivalence Principle with the multi-band timing of the Crab Pulsar}


\author{Zhang Yueyang\altaffilmark{1} and Gong Biping\altaffilmark{1}}
\affil{Physics Department, Huazhong University of Science and Technology, Wuhan, China, 430074}
\email{zhangyuey@hust.edu.cn}
\email{bpgong@hust.edu.cn}



\begin{abstract}
Weak Equivalent Principle (WEP) can be tested through the parameterized post-Newtonian parameter $\gamma$, representing the space curvature produced by unit rest mass. The parameter $\gamma$ in turn has been constrained by comparing the arrival times of photons originating in distant transient events, such as gamma-ray bursts, fast radio bursts as well as giant pulses of pulsars. Those measurements normally correspond to an individual burst event with very limited energy bands and signal-to-noise ratio (S/N). In this letter, the discrepancy in the pulse arrival times of the Crab Pulsar between different energy bands is obtained by the phase difference between corresponding pulse profiles. This allows us to compare the pulse arrival times at the largest energy band differences, between radio and optical, radio and X-ray, radio and gamma-ray respectively. As the pulse profiles are generated by phase-folding thousands of individual pulses, the time discrepancies between two energy bands are actually measured from thousands of events at each energy band, which corresponds to much higher S/N. The upper limit of the $\gamma$ discrepancy set by such an extensively-observed and well-modeled source is as follows: $\gamma_{radio}-\gamma_{\gamma-ray}<3.28\times10^{-9}$ at the energy difference of $E_{\gamma-ray}/E_{radio}\sim10^{13}$, $\gamma_{radio}-\gamma_{X-ray}<4.01\times10^{-9}$ at the energy difference of $E_{X-ray}/E_{radio}\sim10^{9}$, $\gamma_{radio}-\gamma_{optical}<2.63\times10^{-9}$ at $E_{optical}/E_{radio}\sim10^{5}$, and $\gamma_{optical}-\gamma_{\gamma-ray}<3.03\times10^{-10}$ at $E_{\gamma-ray}/E_{optical}\sim10^8$. This actually measures the arrival times of freely-falling photons in the gravitational field of the Milky Way with the largest amount of events and with data of the highest S/N, which tests WEP at the energy band differences that has never been reached before.
\end{abstract}


\section{Introduction}\label{sec1}

Weak Equivalence Principle (WEP) is an important foundation of general relativity and many other metric theories of gravity. WEP can be tested through the parameterized post-Newtonian (PPN) parameters, such as the parameter $\gamma$, which is defined as how much space curvature is produced by unit rest mass \citep[see][]{Will06,Will14}. The accuracy of WEP can be obtained by the $\gamma$ discrepancy for particles with different properties,
since any gravity theory satisfying WEP predicts the same $\gamma$ value. The time interval required for particles to traverse a given distance is longer in the presence of a gravitational potential $U(\mathbf{r})$ by
\begin{equation}\label{shapiro-delay}
\delta t=\frac{1+\gamma}{c^3} |\int_{\mathbf{r}_o}^{\mathbf{r}_e} U(\mathbf{r}) d\mathbf{r}|,
\end{equation}
where $\mathbf{r}_o$ and $\mathbf{r}_e$ are locations of the observer and the emission site of the particles respectively \citep{Shapiro64,Krauss88,Longo88}. So the $\gamma$ discrepancy can be constrained by the time delay between different particles emitted from an astronomical source.

In previous works, the time delay data used to test WEP are all extragalactic transient sources such as
gamma-ray bursts \citep[GRB, e.g.][]{Gao15,Sang16}, fast radio bursts \citep[FRB, e.g.][]{Wei15,Tingay16} and TeV blazers \citep[e.g.][]{Wei16}, with the exception of a 0.4-nanosecond giant pulse of the Crab Pulsar that was used by a recent work of \citet{Yang16}. The time delay between photons of different energies of GRBs and TeV blazers is determined by cross-correlation between light curves obtained from observations at different energy bands \citep{Gao15,Wei16,Nusser16}. In other works, the duration time of an abrupt burst event (a FRB, a short GRB or a giant pulse of a pulsar) is used as a representative of the time delay between the highest and lowest energies within the bandpass of the observing telescope \citep{Wei15,Sang16,Yang16,Tingay16,Nusser16}.
The gravitational potential in consideration is either a Keplerian potential of the Milky Way for sources not too far away, or a cosmological form of the large-scale structure for sources at redshift $z\gtrsim0.5$ \citep{Nusser16,Tingay16}.

In this work, the pulse timing of the Crab Pulsar, a well-studied source with extensive observations at the widest energy coverage, is used to test WEP. And a more realistic gravitational potential of the Milky Way is adopted.

The Crab Pulsar has been extensively observed in energy bands from radio to $\gamma$-ray. The pulse profile of the Crab Pulsar is generated by phase-folding thousands of individual pulses, reaching a very high signal-to-noise ratio (S/N). It exhibits the feature of two most prominent pulse components which is remarkably similar over almost all the energy bands. The time delays between different bands are very small, which are less than $\sim300~\mu$s, corresponding to $\sim1\%$ of the spin period \citep[e.g.][]{Kuiper03,Oosterbroek06}. It is thus believed that the similar pulse profile over all energy bands originates in the same emission region. The accurate time delay measurements based on high-S/N data, the well-modeled pulse profile, and the wide energy coverage of the Crab Pulsar's emission makes it a perfect source to test WEP.

Unlike extragalactic sources, the Crab Pulsar locates in the Milky Way and is very close to the Galactic disc. The gravitational potential of the Milky Way can not be simply considered as a Keplerian potential, but a more complex form.

A brief description of the method of testing WEP is presented in Section \ref{sec2}. Time delay measurements of the Crab Pulsar in literature are discussed in Section \ref{sec3}. In Section \ref{sec4}, the constraints of the $\gamma$ discrepancy using the timing of the Crab Pulsar are shown and compared with previous works. A summary is given in Section \ref{sec5}.
\section{Method description}\label{sec2}

In principle, the observed time delay between photons of two different energy bands  consists of five terms \citep{Gao15,Wei15,Wei16}:
\begin{equation}\label{tobs}
    \Delta t_{obs} = \Delta t_{int} + \Delta t_{LIV} + \Delta t_{spe} + \Delta t_{DM} + \Delta t_{gra}.
\end{equation}

In Eq. \ref{tobs}, $\Delta t_{int}$ is the intrinsic time delay between two photons,  which is determined by the source's radiation mechanism and other characteristics that affect the time delay of photons with different energies. The term $\Delta t_{LIV}$ is the time delay due to the effect of Lorentz invariance violation via an energy-dependent velocity of light. And $\Delta t_{spe}$ is the time delay caused by special-relativistic effects for photons with non-zero rest masses.

The time delay $\Delta t_{DM}$ stems from the dispersion of the line-of-sight free electrons. It is larger for a photon with lower energy and vanishes as the photon energy becomes infinite.  In the standard data reduction process of pulsar timing like TEMPO2 \citep{Hobbs06}, the times of arrival of a pulsar are converted to the ``infinite'' frequency (which is the so called de-dispersion procedure). Therefore, $\Delta t_{DM}$ is negligible in this work.

From Eq. \ref{shapiro-delay}, the time delay $\Delta t_{gra}$ represents the difference in arrival times between two different photons
originated in the gravitational potential $U(\mathbf{r})$, as
\begin{equation}\label{tgra}
    \Delta t_{gra} = \frac{\gamma_{1}-\gamma_{2}}{c^{3}} \vert \int_{\mathbf{r}_{o}}^{\mathbf{r}_{e}} U(\mathbf{r})d\mathbf{r} \vert,
\end{equation}
where $\gamma_1$ and  $\gamma_2$ are the $\gamma$ values of two photons with different energies.

The timing of the Crab Pulsar shows that the low energy photons arrive later than the high energy photons ($\Delta t_{obs}>0$), which is believed to be mainly due to the intrinsic time delay, so that the relationship of $\Delta t_{int}>0$ also holds. Assuming $\Delta t_{gra}>0$, we have $\gamma_1-\gamma_2>0$ by Eq.~\ref{tgra}.
If we also assume $\Delta t_{LIV}+\Delta t_{spe}>0$, so $\Delta t_{gra}=\Delta t_{obs}-\Delta t_{int}-(\Delta t_{LIV}+\Delta t_{spe})<\Delta t_{obs}$.
And hence,
\begin{equation}\label{tobs3}
    \Delta t_{obs} > \frac{\gamma_{1}-\gamma_{2}}{c^{3}} \vert \int_{\mathbf{r}_o}^{\mathbf{r}_e} U(\mathbf{r})d\mathbf{r} \vert.
\end{equation}
This gives the most conservative constraint of $\gamma_1-\gamma_2$.

We adopt a more realistic gravitational potential form, considering it as a two-component system, including a Miyamoto-Nagai disc \citep{Miyamoto75},
\begin{equation}\label{disc}
    \Phi_{disc}(R,z) = -\frac{GM_d}{\sqrt{R^2+(r_a+\sqrt{Z^2+r_b^2})^2}},
\end{equation}
and a NFW \citep{Navarro96} dark matter halo,
\begin{equation}\label{halo}
    \Phi_{halo}(r) = -\frac{GM_{vir}}{r[log(1+c)-c/(1+c)]} log(1+\frac{r}{r_{vir}}),
\end{equation}
with parameters adopted from \citep{Gomez10}. In Eq.~\ref{disc}, $R$ and $z$ are the radial distance and the height in the cylindrical coordinate system, and $r$ in Eq.~\ref{halo} is the radial distance in the spherical coordinate system, which satisfies $r =\sqrt{R^2+z^2}$. Table~\ref{gomez} lists the values of parameters used in this work.

\begin{deluxetable}{ll}
\tablewidth{0pt}
\tablecaption{Parameters of the gravitational potential used in Eq.~\ref{disc} and Eq.~\ref{halo}, cited from \citet{Gomez10}.\label{gomez}}
\tablehead{\colhead{Disc} & \colhead{Halo}}
\startdata
$M_d=7.5\times10^{10}$ M$_\odot$ & $M_{vir}=9\times10^{11}$ M$_\odot$ \\
$r_a=5.4$ kpc  & $r_{vir}$=250 kpc \\
$r_b=0.3$ kpc  & $c=13.1$ \\
\enddata
\end{deluxetable}

So, the gravitational potential is the sum of the two potentials described by Eq.~\ref{disc} and Eq.~\ref{halo}, as $U(R,z) = \Phi_{disc}(R,z) + \Phi_{halo}(r)$.

Integrating along a straight line from the Sun to the Crab Pulsar, $U(R,z)$ can be characterized as $U(R)$, since $z$ is a function of $R$. Then we have
\begin{equation}
     \gamma_1 - \gamma_2 < \frac{\Delta t_{obs} c^3} {\vert \int_{R_o}^{R_e} U(R) dR \vert},
\end{equation}
where $R_o$ stands for the radial distance of the Sun relative to the Galactic center and $R_e$ is the radial distance of the Crab Pulsar. We adopt $R_o = 8.3$ kpc and $z_o = 15$ pc. The distance from the Crab Pulsar to the Sun is $d=2.0$ kpc \citep{Kaplan08}. The Galactic coordinate of the Crab Pulsar is ($l \approx 184.56^{\circ}, b \approx -5.78^{\circ}$). $R_e$ and $z_e$ can be easily calculated by $R_e = \sqrt{R_o^2+(d\cos b)^2-2 R_o d \cos b \cos(360^{\circ}-l)} \approx 10.3$ kpc and $z_e = z_o + d \sin{b} \approx -186$ pc. The relation between $z$ and $R$ is $z = -0.1005 R + 0.8492$.

Then, the discrepancy between  $\gamma_1$ and $\gamma_2$ corresponding to two photons with different energies relates with their time discrepancy $\Delta t_{obs}$, by,
\begin{equation}\label{dgamma}
     \gamma_1 -\gamma_2 < 1.167\times 10^{-5}~\mbox{s}^{-1}~\Delta t_{obs}.
\end{equation}

Consequently, WEP can be tested by Eq.~\ref{dgamma} with the values of $\Delta t_{obs}$ between various energy bands.

\section{Time delay measurements of the Crab Pulsar}\label{sec3}

The latest time delay measurements of the Crab Pulsar are summarized in Table~\ref{time-delay-list}, for optical, X-ray and $\gamma$-ray pulses in comparison with radio pulses. We also obtain a optical-$\gamma$-ray time delay by comparing data of optical and $\gamma$-ray in this table. Most of the measurements have used the Crab Pulsar Monthly Ephemeris of the Jodrell Bank Observatory \citep{Lyne93} to compare with the pulse arrival times of higher energy bands, except \citet{Oosterbroek08} and \citet{Abdo10} who carried out their own radio observations.

\begin{deluxetable}{lllllll}
\tabletypesize{\tiny}
\rotate
\tablecaption{The time delays of the Crab Pulsar for optical, X-ray and $\gamma$-ray pulses in comparison with radio pulses.\label{time-delay-list}}
\tablewidth{0pt}
\tablehead{\colhead{radio - band} & \colhead{time delay} & \colhead{origin of radio data} & \colhead{instrument (band)} & \colhead{instrument (radio)} & \colhead{DM uncertainty} & \colhead{Reference} \\
\colhead{}                        & \colhead{($\mu$s)}   & \colhead{}                     & \colhead{}                  & \colhead{}                   & \colhead{(pc cm$^{-3}$)}          & \colhead{}    }
\startdata
radio - optical      & 273$\pm100$ & CPME\tablenotemark{2} & WHT\tablenotemark{3} and OGS\tablenotemark{4} & JBO\tablenotemark{10} & 0.005 & \citet{Oosterbroek06} \\
             & 255$\pm21$\tablenotemark{1} & observation & OGS               & Nan\c{c}ay radio telescope & 0.005 & \citet{Oosterbroek08} \\
             & 230$\pm60$ & CPME & Copernico Telescope & JBO & 0.005 & \citet{Germana12} \\
radio - X-ray        & 280$\pm45$ & CPME & INTEGRAL\tablenotemark{5}~, RXTE PCA\tablenotemark{6}      & JBO & 0.005 & \citet{Kuiper03} \\
             & 344$\pm40$\tablenotemark{1} & CPME & RXTE PCA  & JBO & 0.005 & \citet{Rots04} \\
             & 275$\pm43$ & CPME & INTEGRAL & JBO & 0.005 & \cite{Molkov10} \\
             & 306$\pm53$ & CPME & XMM-Newton\tablenotemark{7} & JBO & 0.005 & \citet{Martin-Carrillo12} \\
radio - $\gamma$-ray & 241$\pm104$ & CPME & EGRET\tablenotemark{8} & JBO & 0.005 & \citet{Kuiper03} \\
             & 281$\pm24$\tablenotemark{1}  & observation & Fermi LAT\tablenotemark{9} & Nan\c{c}ay radio telescope & 0.0003 & \citet{Abdo10} \\
\tableline
optical - $\gamma$-ray & 26$\pm32$\tablenotemark{1} & \multicolumn{5}{l}{calculated by $(281\pm24)-(255\pm21)$}\\
\enddata
\tablenotetext{1}{The data marked with ``1'' are used in the test of this letter.}
\tablenotetext{2}{Crab Pulsar Monthly Ephemeris}
\tablenotetext{3}{William Hershel Telescope}
\tablenotetext{4}{Optical Ground Station Telescope}
\tablenotetext{5}{International Gamma-Ray Astrophysics Laboratory}
\tablenotetext{6}{Proportional Counter Array on the Rossi X-ray Timing Explorer}
\tablenotetext{7}{X-ray Multi-Mirror Mission}
\tablenotetext{8}{Energetic Gamma Ray Experiment Telescope}
\tablenotetext{9}{Fermi Large Area Telescope}
\tablenotetext{10}{Jodrell Bank Observatory}
\end{deluxetable}

Unlike GRBs and TeV blazers, the time delays of which are determined by cross-correlation between light curves at different energy bands, the arrival times of a pulsar can be directly measured because of the prominent peak displayed in the pulse profile. The pulse arrival time of the Crab Pulsar at each energy band is defined by the phase of the main pulse peak. And the time delay between two energy bands is calculated from the phase discrepancy between them.

The highly stable profile of a pulsars allows one to phase-fold thousands of individual pulses, so that the integrated pulse profile obtained is of a very high S/N. This is analogous to adding up thousands of identical transient events to obtain the overall light curve. Taking \citet{Abdo10} for example, the Crab Pulsar was observed in radio band by Nan\c{c}ay radio telescope with an integration time of 1 minute, which means the integrated pulse profile is generated by phase-folding $\sim1800$ periods, resulting in an S/N of $\gtrsim1000$ \cite[see Figure 1 of][]{Abdo10}. The $\gamma$-ray emission of the Crab Pulsar is recorded as isolated photon events on the pulsar's coordinates in a time sequence. $14,563\pm240$ pulsed $\gamma$-ray photons were obtained from 248 days of data. Based on the radio ephemeris, the times of photon events are converted into phases within the pulsar period. In other words, the pulse profile of $\gamma$-ray emission is generated by phase-folding data of 248 days. The S/N of the $\gamma$-ray profile is still $\gtrsim100$ in spite of much lower $\gamma$-ray flux intensity.

The uncertainty of the time delay measurements is usually separated into two parts, the statistical and systematic uncertainty respectively. The former comes from the procedure of fitting the pulse profiles of the two bands in comparison by a template of the profile, which aims to accurately determine the peak of the pulse profile. The latter one stems from the error of clocks used by the instruments and the error of the dispersion measure obtained from the radio observation. Here the two terms are added in quadrature.

We selected the radio-optical and radio-$\gamma$-ray time delays with the smallest uncertainties in the test, i.e. $\Delta t_{radio-optical}=255\pm21~\mu$s and $\Delta t_{radio-\gamma-ray}=281\pm24~\mu$s. The measurement of \cite{Rots04} $\Delta t_{radio-X-ray}=344\pm40~\mu$s deviates from the rest of radio-X-ray data considerably. In order to make a conservative calculation, it is still selected because it corresponds to the largest discrepancy between the radio and X-ray data. The optical-$\gamma$-ray time delay $\Delta t_{optical-\gamma-ray}=26\pm32~\mu$s is also included in the test.

\section{Result and discussion}\label{sec4}

Consequently, WEP can be tested via Equation~\ref{dgamma} by the discrepancies of $\gamma$ values between different energy bands:
\begin{equation}\label{dgamma-radio-optical}
    \gamma_{radio} - \gamma_{optical} < 2.63\times 10^{-9},
\end{equation}
\begin{equation}\label{dgamma-radio-Xray}
    \gamma_{radio} - \gamma_{X-ray} < 4.01\times 10^{-9},
\end{equation}
\begin{equation}\label{dgamma-radio-gamma}
    \gamma_{radio} - \gamma_{\gamma-ray} < 3.28\times 10^{-9},
\end{equation}
\begin{equation}\label{dgamma-optical-gamma}
    \gamma_{optical} - \gamma_{\gamma-ray} < 3.03\times10^{-10},
\end{equation}

As a comparison, the new constraints of the $\gamma$ discrepancy (Eq. \ref{dgamma-radio-optical}-\ref{dgamma-optical-gamma}) and some of the best previous results are listed in Table \ref{dgamma-list}. The parameter $E_{high}/E_{low}$ is defined to indicate the energy band difference in each comparison, where $E_{high}$ and $E_{low}$ are the higher and lower energy bands respectively.

\begin{table*}\tiny
\caption{The upper limits of the $\gamma$ discrepancy given by this work and previous works }\label{dgamma-list}
\begin{tabular}{lllll}
\tableline
\tableline
Source name & Test particles and energy bands & Upper limit of $\Delta\gamma$ & $E_{high}/E_{low}$ & Reference \\
\tableline
Crab Pulsar & photon(radio) - photon(optical) & $2.63\times10^{-9}$ & $\sim10^5$ & Eq. \ref{dgamma-radio-optical} \\
Crab Pulsar & photon(radio) - photon(X-ray) & $4.01\times10^{-9}$ & $\sim10^9$ & Eq. \ref{dgamma-radio-Xray} \\
Crab Pulsar & photon(radio) - photon($\gamma$-ray) & $3.28\times10^{-9}$ & $\sim10^{13}$  & Eq. \ref{dgamma-radio-gamma} \\
Crab Pulsar & photon(optical) - photon($\gamma$-ray) & $3.03\times10^{-10}$ & $\sim10^8$ & Eq. \ref{dgamma-optical-gamma} \\
\tableline
GRB 080319B & photon(eV) - photon(MeV) & $2.3\times10^{-10}~(3\sigma)$ & $10^6$ & \citet{Nusser16} \\
            &                          & $1.3\times10^{-11}~(2\sigma)$ &        &                 \\
Crab Pulsar (giant pulse) & photon(8.15 GHz) - photon(10.35 GHz) & (0.6-1.8)$\times10^{-15}$ & $\sim1.2$ & \citet{Yang16} \\
\tableline
\end{tabular}
\end{table*}

The first three constraints Eq. \ref{dgamma-radio-optical}-\ref{dgamma-radio-gamma} test WEP by comparing radio pulses with those of higher energy bands. WEP has never been tested with these energy band differences before. Moreover, the radio-X-ray and radio-$\gamma$-ray constraints have the largest energy band differences of $10^9$ and $10^{13}$ respectively, with 3 to 7 orders larger than those of previous works.

In previous works, the test with the largest energy band difference and the relatively most stringent constraint is from \citet{Nusser16}, which gives $\gamma_{eV}-\gamma_{MeV}<2.3\times10^{-10}~(3\sigma)$, or $\gamma_{eV}-\gamma_{MeV}<1.3\times10^{-11}~(2\sigma)$. In contrast, Eq. \ref{dgamma-optical-gamma} sets a similarly stringent constraint at the similar energy range, but with an energy band difference 2 orders larger than \citet{Nusser16}.

In general, the new results extend the test of WEP to the largest energy band differences. And in the case of comparable constraints of the $\gamma$ discrepancy at the optical-$\gamma$-ray comparison, the energy band difference of the new result is 2 orders larger than \citet{Nusser16}.

Up to date, the most stringent constraint of the $\gamma$ discrepancy is $\gamma(8.15~\mbox{GHz})-\gamma(10.35~\mbox{GHz})<(0.6-1.8)\times10^{-15}$, obtained from a 0.4-nanosecond giant pulse duration time of the Crab Pulsar by \cite{Yang16}, but it only covers a very narrow radio frequency band of 2.2 GHz.

\section{Summary}\label{sec5}

The accuracy of WEP can be characterized by the discrepancy in the parameter $\gamma$, for photons with different energies. Unlike the transient signals such as GRBs and giant pulses of pulsars, a non-transient signal, the timing of the Crab Pulsar with high-S/N data, is applied in the test of WEP at multiple energy bands from radio to $\gamma$-ray. The test of WEP is thus extended to energy band differences up to $E_{high}/E_{low}\sim10^{13}$ which has never been achieved before. The new constraint for the optical-$\gamma$-ray test is comparable to the best previous result but with a 2-order larger energy band difference.

Comparing with previous works, the new method has three advantages:
\begin{itemize}
  \item A more realistic gravitational potential of the Milky Way is applied when calculating the time delay $\Delta t_{gra}$, which makes the constraint more accurate.
  \item The pulsation of the Crab Pulsar is stable, periodic, and always available for observation. The high-S/N pulse profile obtained by the phase-folding technique makes the time delay data of the pulsar can be measured with high precisions.
  \item If WEP was not satisfied by a gravity theory, the parameter $\gamma$ would be different for photons with different energies. Then the photons with the most different energies would be most likely to reveal the $\gamma$ discrepancy. The multi-band pulsation of the Crab Pulsar from radio to $\gamma$-ray provides a great opportunity to test WEP at the largest energy band difference.
\end{itemize}

The intrinsic time delay of the Crab Pulsar is most likely due to the different locations of the emission sites of different energy bands, as the pulse periods of different energy bands are identical. Under such a circumstance, a time delay of $\sim 300~\mu$s can be explained by two reasons: firstly the emission region of radio band locates at an emission region 90 km lower than that of the higher band, and secondly between the radio beam and that of the higher band exists an angle of $\sim3.3^{\circ}$. The detailed structure and geometry of the emission region of the Crab Pulsar will allow us to get more precise values of $\Delta t_{int}$. With more understanding of the emission region of the Crab Pulsar, more stringent limits on WEP are expected in the future.

\acknowledgments
We wish to thank Dick Manchester and George Hobbs for helpful discussion and suggestions. And we thank Hao Tong and Mingyu Ge for communication on the latest progress on the time delay measurements of the Crab Pulsar. This research is supported by the National Natural Science Foundation of China, under the grant NSFC11373018 and NSFC11503006. We also thank for the support of the grant Beyond the Horizons 2012.

%% the file, uncomment the \plotone, \plottwo, and
%% If you need a layout that cannot be achieved with \plotone or
%% \plottwo, you can invoke the graphicx package directly with the
%% \includegraphics command or use \plotfiddle. For more information,
%% please see the tutorial on "Using Electronic Art with AASTeX" in the
%% documentation section at the AASTeX Web site,
%% http://www.journals.uchicago.edu/AAS/AASTeX.
%%
%% The examples below also include sample markup for submission of
%% supplemental electronic materials. As always, be sure to check
%% the instructions to authors for the journal you are submitting to
%% for specific submissions guidelines as they vary from

%% This example uses \plotone to include an EPS file scaled to
%% 80% of its natural size with \epsscale. Its caption
%% has been written to indicate that additional figure parts will be
%% available in the electronic journal.

%\begin{figure}
%\epsscale{.80}
%\plotone{f1.eps}
%\caption{Derived spectra for 3C138 \citep[see][]{heiles03}. Plots for all sources are available
%in the electronic edition of {\it The Astrophysical Journal}.\label{fig1}}
%\end{figure}

\clearpage

%% The following command ends your manuscript. LaTeX will ignore any text

\end{document}